# Reinforcement Learning Based Gasoline Blending Optimization: Achieving More Efficient Nonlinear Online Blending of Fuels

**Muyi Huang[a], Renchu He[a, *], Xin Dai[a], Xin Peng[a], Wenli Du[a,b], Feng Qian[a, b, *]**

[a] Key Laboratory of Smart Manufacturing in Energy Chemical Process, Ministry of Education, East China University of Science and Technology, Shanghai, China

[b] Shanghai Institute of Intelligent Science and Technology, Tongji University, Shanghai, China

**Abstract:** The online optimization of gasoline blending benefits refinery economies. However, the nonlinear blending mechanism, the oil property fluctuations, and the blending model mismatch bring difficulties to the optimization. To solve the above issues, this paper proposes a novel online optimization method based on deep reinforcement learning algorithm (DRL). The Markov decision process (MDP) expression are given considering a practical gasoline blending system. Then, the environment simulator of gasoline blending process is established based on the MDP expression and the one-year measurement data of a real-world refinery. The soft actor-critic (SAC) DRL algorithm is applied to improve the DRL agent policy by using the data obtained from the interaction between DRL agent and environment simulator. Compared with a traditional method, the proposed method has better economic performance. Meanwhile, it is more robust under property fluctuations and component oil switching. Furthermore, the proposed method maintains performance by automatically adapting to system drift.

---

# 1. Introduction

Gasoline is an important product for refineries. The production of gasoline generates 60%-70% of the refinery's total profits[1]. Gasoline blending is a critical part of the gasoline production process. Thus, the gasoline blending optimization provides substantial economic benefits to the refinery.

The gasoline blending optimization system consists of three levels: offline optimizer or scheduler, online optimizer, and regulatory control [2, 3]. At first level, the offline optimizer or scheduler uses the offline feedstock data to develop an initial blending recipe for gasoline blending over the next few days. Then the initial blending recipe is downloaded to the online optimizer in the second level. The online optimizer uses online feedstock data to optimize the initial blending recipe during the blending process and submits the optimized final blending recipe to the regulatory control level. As one of the critical parts of the gasoline blending optimization system, online optimizers need to consider two aspects. One is to minimize the gap between the initial and final recipe and the other is to reduce the cost of the final recipe. At the same time, properties of the blended oils based on the recipe need to meet certain limits. The properties include research octane number (RON), motor octane number (MON), and Reid vapor pressure (RVP), etc.

As one of the commonly used methods for online optimization, linear programming algorithms [4] neglect the nonlinear features of gasoline blending. For example, the crucial gasoline properties, such as RON and RVP, are blended in a nonlinear mechanism. As a result, nonlinear programming methods based on the model of crucial properties, including ethyl RT-70 [5] and Stewart model [6] for RON and Chevron model for RVP [7], are applied in the online optimization. However, most of them require a large amount of experimental data to determine the parameters in the models. In addition, the model mismatch caused by the fluctuation of component oil properties or plant/model parametric mismatch may directly lead to the failure of the optimal recipe. For example, the optimal recipe derived from the model may not guarantee that the crucial properties satisfy the constraints.[8]. The traditional solution to model mismatch is deviation compensation, among which the blending effect model [9] based optimization method has remarkable adaptability in field applications. This method has a simple linear structure and efficiently utilizes production data. However, the study [10] points out that even if the deviation compensation method is adopted, in the case of component oil property fluctuation, the linear or nonlinear optimization method cannot ensure the solution's optimality. Further, since gasoline is blended in batches, more advanced methods that consider the entire blending process rather than the situation of single optimization period are proposed [11, 12]. However, the high time cost of online calculations will make practical application difficult because a nonlinear optimization problem considering the whole-time horizon of blending has to be solved. Therefore, the nonlinearity of the blending mechanism, the determination and updating of blending model parameters, the fluctuation of component oil properties and the contradiction between global optimum and computation time are the main issues to be resolved in the online blending process.

Different from the traditional methods aforementioned, reinforcement learning (RL) is promising to deal with the above issues. When using RL methods, the problem needs to be first formalized as a Markov Decision Process (MDP) [14]. As illustrated in Fig.1, RL involves an agent which takes the action $A_t$ based on the state $S_t$ and the reward $R_t$. The stated is used to represent the feature of the environment and the reward is used to evaluate the action. Then the environment receives the action returns the new state $S_{t+1}$ and the new reward $R_{t+1}$. During this interaction, the agent learns a policy to map states to actions by maximizing the cumulative rewards. RL methods avoid the difficulty of modeling the process because they learn the policy directly from the data, and their unique learning ability enables the RL agent to follow the drift of the environment [13]. Therefore, when RL methods are adopted, we no longer need to model the nonlinearity in the blending mechanism. By constantly obtaining data from the blending process and updating the policy, RL methods can perceive the property fluctuation of component oil and automatically adjust model parameters to follow the blending system's system drift in the actual operation process. Since RL is a learning mechanism to obtain the maximum cumulative rewards, a RL agent considers the impact of a decision on subsequent processes each time the decision is made. Thus RL methods have the ability to consider the blending problem in the whole time horizon. Further, a fully trained RL agent can produce blend recipes rapidly to be used for online optimization.

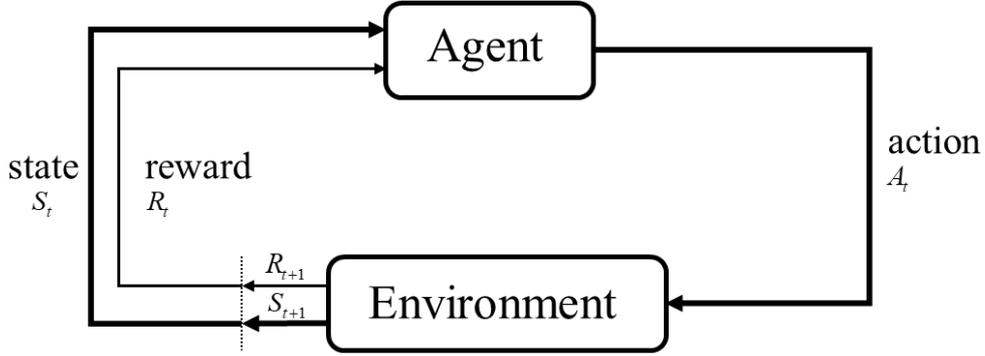

Fig. 1. Interaction between agent and environment in MDP

According to the size of the problem to be solved, the algorithms in traditional reinforcement learning can be divided into two categories The first type is the tabular solution method. This type of method is often used when the state and action space of the problem is small enough to be represented by a table. There are three basic methods for this kind of problem: dynamic programming [14], the Monte Carlo method [15], and the TD method [16]. The second type is approximate solution methods, which are mainly used to deal with problems with large state and action spaces. This type of method usually uses a parameterized function to model the policy. Typical methods include the qualification trace method [17] and the policy gradient method (PG) [18], etc.

However, real-world problems are often more complex. The state and action spaces of these problems are large and continuous. Moreover, it is also required that the agent learns better and more generalized policies. It is difficult to solve such problems with traditional RL algorithms in this situation. In recent years, deep reinforcement learning algorithm (DRL) was developed by DeepMind based on the combination of reinforcement learning and deep learning. The proposed Deep Q network (DQN) [19] is used to play Atari games, which has achieved competitive results not inferior to human players. The emergence of DRL makes it possible to solve complex problems.

With the development of research, a large number of DRL algorithms have been proposed. Depending on how the policy is updated, DRL algorithms can also be divided into two categories: on-policy and off-policy. On-policy methods include TRPO [20], PPO [21], A3C [22], etc. The data used in updating their policy is restricted. Only data generated by the policy itself can be used. Consequently, whenever an old policy is updated to a new policy, the data used for this update process is discarded, since updating the new policy cannot use the old data. As a result, new data are required to be collected for each decision step, leading to poor sample efficiency. While in off-policy methods, such as Deep Q Network (DQN) [19], Deep Deterministic Policy Gradient (DDPG) [23], or Twin Delayed Deep Deterministic Policy Gradient (TD3) [24], old data of the refinery can also be used in the process of policy update. In this way, data utilization is greatly improved. However, such methods tend to have poor stability and convergence during training process [25].

In recent years, DRL has found many applications in the process optimization and control. Oh et al. proposed an actor-critic reinforcement learning optimization strategy using a DNN surrogate model for determining the optimal operating conditions for hydrocracking units [26]. Joshi et al. proposed Twin actor twin delayed deep deterministic policy gradient RL controller for batch process , and designed two new reward functions for the controller [27]. Goulart et al. developed an autonomous pH controller for electroplating industry liquid effluents, based on fully automated Reinforcement Learning [28]. Heidari et al. developed a DRL-based control framework to integrate the occupants' behavior into hot water systems control, which can balance water hygiene, comfort, and energy use [29]. Zeng et al. adopted the DRL method to solve the problem of large-scale optimization of heliostat field aiming strategy, which provided better, or comparable performance compared to heuristic optimization methods with an order of magnitude less computation time [30]. Despite the growing research and applications of DRL, no relevant DRL method is available for online optimization of gasoline blending, to the best of our knowledge. As mentioned above, the characteristics of DRL can solve the issues in online optimization of gasoline blending, which makes the application of DRL meaningful.

In this work, a DRL based online optimization method is proposed for gasoline blending. We first establish a MDP model for the gasoline blending process. Then, SAC, a state-of-art DRL algorithm, is introduced for deciding blending recipes at each optimization period. The blending recipes are then sent to the simulator representing the actual blending process. The simulator gives rewards according to the adjustment of the recipes and then returns the information required for the next optimization period to the DRL agent. In order to simulate the complex situation of the actual production environment, one-year measurement data from a real-word refinery is used to extract the fluctuating properties of the component oils. Besides, the scenario of component oil depletion is also taken into account. Data during this interaction is collected in a reply buffer which helps the DRL agent learn the reasonable policy. The trained agent is compared with the optimization method based on a blending effect model to indicate the effectiveness of the proposed method. In addition, the adaptability of the proposed method in the face of system drift is detected. The main contributions of this study are listed as follows.

(1) A form of MDP expression is established for the gasoline blending problem, which considers the blending cost, the properties of the component oils, and the property constraints in the mixed oil tube and the blending tank.

(2) A DRL based online optimization method is proposed to avoid the modeling difficulties in gasoline blending and follow the system drift. The proposed method considers the whole time-horizon of the blending process.

(3) A practical case study of the gasoline blending system is studied to demonstrate the advantages of the proposed method. The performance of the proposed method is compared with that of the optimization method based on the blending effect model.

The rest of the article is organized as follows. The process of the gasoline blending system is introduced in Section 2. . The online optimization method based on DRL is described in Section 3. A practical case study is conducted in Section 4 to analyze the performance of the proposed DRL based optimization method. The conclusions of this study are given in Section 5.

## 2. Problem statement

Commonly, gasoline blending processes can be classified into two categories: tank blending and pipeline blending [31]. In this work, we consider the pipeline blending process, the structure of which is shown in Fig.2. Various automation equipment is used to control the flow of each component oil participating in the blending. The component oils will be mixed according to the recipe at the static mixer through the pipeline. Then the mixed oil will be transported to the storage tank. Properties of component oils and mixed oils can be obtained by a near-infrared spectrometer. The properties of the oils are measured every 30 mins by the near-infrared probes on each component oil pipeline and the mixed oil pipeline.

In the actual blending process, gasoline storage tanks often contain bottom oil, the properties of which differ significantly from the standards for the product oil. It is necessary to gradually correct the properties of the bottom oil during the blending process to reach the standard of the product oil.

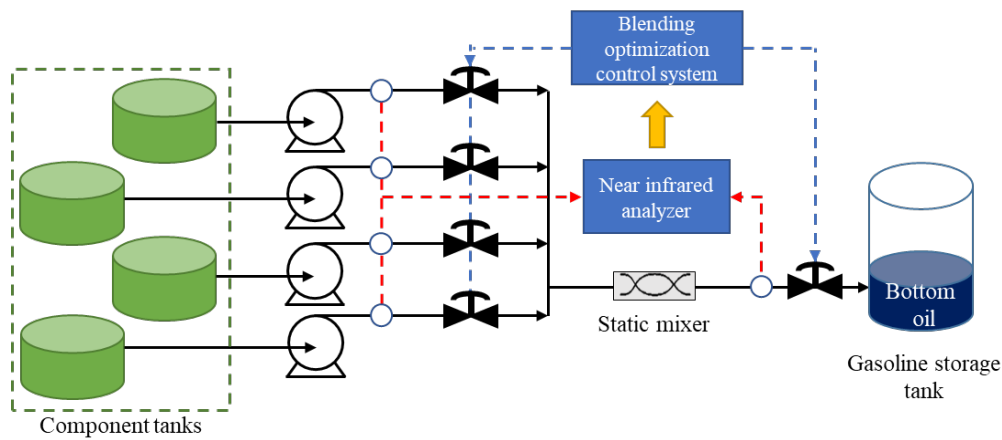

Fig. 2. Gasoline blending process.

In this blending process, the difficulty lies in dealing with the nonlinearity of the blending mechanism and the fluctuating component oil properties. At the same time, it is necessary to continuously update the model to adapt to the system drift in the actual process.

The blending optimization system aims to produce qualified gasoline at the least cost under the condition of meeting process constraints. Therefore, the gasoline blending problem is formalized as in Eq.(1)

$$\min \ Cost_t = \sum_{i=1}^{n} Ccompo_i \cdot DevRecipe_{i,t}$$

s.$t$

$$DevRecipe_{\min} \leq DevRecipe_{i,t} \leq DevRecipe_{\max}, \forall i,t \quad (1)$$
$$DevInitRecipe_{\min} \leq DevInitRecipe_{i,t} \leq DevInitRecipe_{\max}, \forall i,t$$
$$PLoBound_{j,t} \leq Pmix_{j,t} \leq PUpBound_{j,t}$$

where $Cost_t$ is the reward function with respect to the cost of mixed oil. $DevRecipe_{i,t}$ is the change of the component oil recipe from the last optimized recipe and $DevInitRecipe_{i,t}$ is the change of the component oil recipe from the initial recipe. $Pmix_{j,t}$ is the property of the mixed oil.

## 3. DRL based online gasoline blending optimization

### 3.1 Reinforcement learning algorithm

RL is a machine learning method based on the MDP framework to solve sequential decision-making problems. In this work, there is a learning agent that can constantly ($t \in [1, N]$) interact with the gasoline blending environment. The agent gives the blending recipes $a_t$ based on the observed state $s_t$ such as the properties of component oil and property constraints of mixed oil, and receives the associated reward $r_t$. The reward represents the quality of the given recipes. Then, the gasoline blending environment transfers the current state $s_t$ to $s_{t+1}$ according to the blending recipes $a_t$ and characteristics of the environment itself, and outputs a new reward $r_{t+1}$. The main purpose of the reward is to help the agent give a reasonable recipe by rewarding the low-cost recipes and penalizing ones against constraints. The learned probability distribution $\pi$ is called a policy for action sampling based on the observed state. The goal of the learning process is to maximize the sum of the future rewards and find the optimal policy $\pi^*$ :

$$\pi^* = \arg\max_{\pi} E_{(s_t,a_t)\sim \rho_\pi}[\sum_t \gamma^t r(s_t,a_t)] \qquad (2)$$

where $\rho_\pi$ denotes the distribution of the state-action pairs under the policy $\pi$. $\gamma$ is a discount factor where $0 \leq \gamma \leq 1$, which makes immediate rewards more valuable than future rewards.

In this work, the soft actor-critic (SAC) algorithm [32] is implemented. SAC is based on the actor-critic framework, which means that there are two types of networks in the algorithm: actor and critic. The actor is responsible for performing an action, and the critic is responsible for giving scores according to the state and action. As an off-policy DRL algorithm, SAC has high data efficiency, but it also has the characteristics of stable training of on-policy DRL algorithms. SAC considers entropy $H(\pi(s_t))$ as a part of the reward function. This encourages the agent to explore the optimal policy in different ways, so such agent is more adaptable to disturbance and the stability of convergence is improved:

$$\pi^* = \arg\max_{\pi} E_{(s_t,a_t)\sim \pi}[\sum_t \gamma^t (r(s_t,a_t) + \alpha H(\pi(s_t)))] \qquad (3)$$

where $\alpha$ is the temperature parameter which balances the return and the entropy.

As shown in Fig.3, the overall framework of SAC consists of five parts: (1) The data buffer $D$. Each time the agent interacts with the environment, a data tuple $s_t, a_t, r_t, s_{t+1}$ is generated. Through constant interactions, a large number of data tuples are collected into the data buffer $D$. When the number of data tuples exceeds the storage limit of the data buffer $D$, the old data tuples will be replaced by the new ones to follow the latest changes in the system. (2) State value function $V_\psi(s_t)$, which represents the value of the given state $s_t$. (3) Target state value function $V_{\tilde{\psi}}(s_t)$, where $\tilde{\psi}$ can be an exponentially moving average of the value network weights.

Target value networks can be used to stabilize the training process. (4) Policy function $\pi_\varphi(a_t|s_t)$, which provides the distribution of action according to the given state $s_t$. (5) Soft Q-function $Q_\theta(s_t,a_t)$. $Q_\theta(s_t,a_t)$ is used to estimate the soft value of taking action $a_t$ in state $s_t$. Except for the first part, the rest can be modeled by neural network, while the parameters of these networks are $\psi, \tilde{\psi}, \varphi$ and $\theta$.

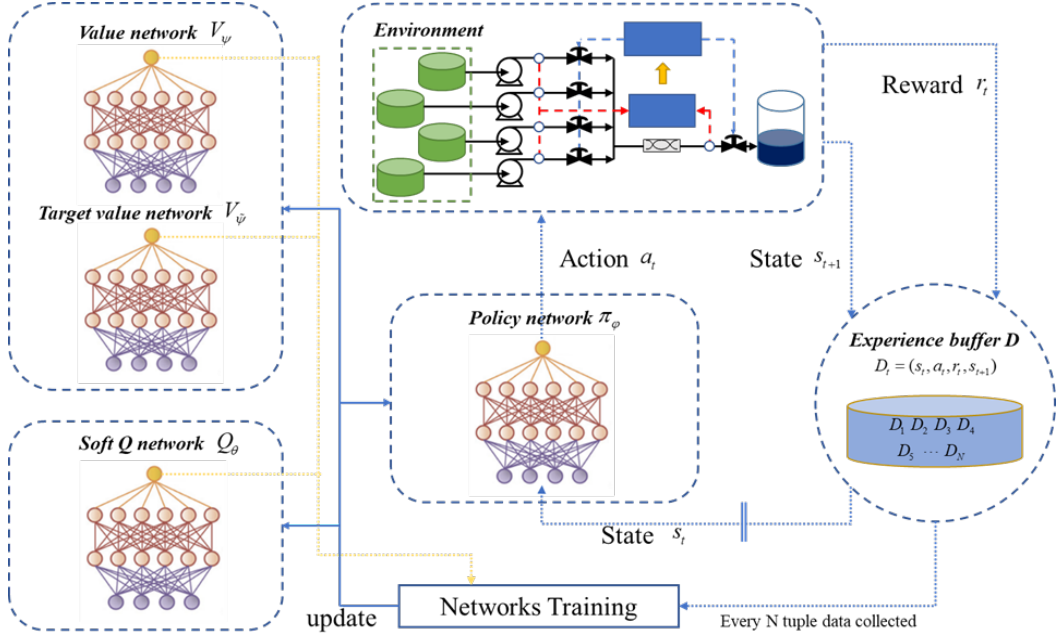

Fig. 3. The framework of SAC.

The pseudo-code for the SAC algorithm is given in Algorithm 1. The equations for optimizing the parameters $\psi, \theta, \phi$ can be expressed by Eqs.(4)-(6), respectively.

$$\hat{\nabla}_\psi J_V(\psi) = \nabla_\psi V_\psi(s_t)(V_\psi(s_t) - Q_\theta(s_t,a_t) + \log \pi_\phi(a_t|s_t)) \qquad (4)$$

$$\hat{\nabla}_\theta J_Q(\theta) = \nabla_\theta Q_\theta(a_t,s_t)(Q_\theta(s_t,a_t) - r(s_t,a_t) - \gamma V_{\bar{\psi}}(s_{t+1})) \qquad (5)$$

$$\hat{\nabla}_\phi J_\pi(\phi) = \nabla_\phi \log \pi_\phi(a_t|s_t) \\ + (\nabla_{a_t} \log \pi_\phi(a_t|s_t) - \nabla_{a_t} Q(s_t,a_t))\nabla_\phi a_t \qquad (6)$$

where

$$a_t = f_\phi(\varepsilon_t;s_t) \qquad (7)$$

where $\varepsilon_t$ is an input noise vector sampled from spherical Gaussian.

**Algorithm 1** Soft Actor-Critic
---
Initialize parameter vectors $\psi, \tilde{\psi}, \theta, \phi$.

**for** each iteration **do**

    **for** each environment step **do**

        $a_t \sim \pi_\phi(a_t | s_t)$

        $s_{t+1} \sim p(s_{t+1} | s_t, a_t)$

        $D \leftarrow D \cup \{(s_t, a_t, r(s_t, a_t), s_{t+1})\}$

    **end for**

    **for** each gradient step **do**

        $\psi \leftarrow \psi - \lambda_V \hat{\nabla}_\psi J_V(\psi)$

        $\theta_i \leftarrow \theta_i - \lambda_Q \hat{\nabla}_\theta J_Q(\theta_i)$ for $i \in \{1, 2\}$

        $\phi \leftarrow \phi - \lambda_\pi \hat{\nabla}_\phi J_\pi(\phi)$

        $\tilde{\psi} \leftarrow \tau\psi + (1-\tau)\tilde{\psi}$

    **end for**

**end for**

---

## 3.2 MDP expression for online gasoline blending

In a gasoline blending system with $n$ component oils and $m$ properties, the state at moment $t$ is listed as follows:

1. Component oil property matrix:

$$Pcompo_t = [Pcompo_{1,t}\ Pcompo_{2,t}\ \cdots\ Pcompo_{n,t}]^T,$$

    where $Pcompo_{i,t} = [Pcompo_{i,1,t}\ Pcompo_{i,2,t}\ \cdots\ Pcompo_{i,m,t}]^T$.

2. Component oil recipe vector: $Recipe_t = [Recipe_{1,t}\ Recipe_{2,t}\ \cdots\ Recipe_{n,t}]^T$.

3. Component oil recipe deviation vector:

$$DevInitRecipe_t = [DevInitRecipe_{1,t}\ DevInitRecipe_{2,t}\ \cdots\ DevInitRecipe_{n,t}]^T.$$

4. Mix oil cost: $Cmix_t = \sum_{i=1}^{n} Ccompo_i \cdot Recipe_{i,t}$.

5. Mix oil property vector: $Pmix_t = [Pmix_{1,t}\ Pmix_{2,t}\ \cdots\ Pmix_{m,t}]^T$.

The component oil property matrix and the mix oil property vector can be formed by the measurement through near-infrared spectroscopy.

6. Property vector of oil in the gasoline storage tank:

$$Ptank_t = [Ptank_{1,t} \ Ptank_{2,t} \ \cdots \ Ptank_{m,t}]^T.$$

Property of oil in the gasoline storage tank usually cannot be measured but can be estimated by linear summation of the property of the mixed oil to be input into the tank and the property of existing oil in the tank according to their volume ratio.

7. The volume of oil in gasoline storage tank: $Vtank_t$.

8. Mix oil property lower and upper constraint vector:

$$PLoBound_t = [PLoBound_{1,t} \ PLoBound_{2,t} \ \cdots PLoBound_{m,t}]^T$$

$$PUpBound_t = [PUpBound_{1,t} \ PUpBound_{2,t} \ \cdots PUpBound_{m,t}]^T$$

The property bound constantly change in the blending process for it should be corrected according to the deviation between the property of the oil in the storage tank and the gasoline production property standard, which can be calculated as follows:

$$PLoBound_{j,t} = \frac{PLoPO_j \cdot V_{target} - Ptank_{j,t} \cdot Vtank_t}{Vtarget - Vtank_t} \quad (8)$$

$$PUPBound_{j,t} = \frac{PUpPO_j \cdot V_{target} - Ptank_{j,t} \cdot Vtank_t}{Vtarget - Vtank_t} \quad (9)$$

The action is a vector:

$DevRecipe_t = [DevRecipe_{1,t} \ DevRecipe_{2,t} \ \cdots \ DevRecipe_{n-1,t}]^T$, which corresponds to the recipe change of each component oil. The dimension of the vector is n-1 because the recipe change should meet the following constraint:

$$\sum_{i=1}^{n} DevRecipe_{i,t} = 0 \quad (10)$$

The $Recipe_{i,t}$ and $DevInitRecipe_{i,t}$ are described as follows:

$$Recipe_{i,t+1} = Recipe_{i,t} + DevRecipe_{i,t} \quad (11)$$

$$DevInitRecipe_{i,t+1} = DevInitRecipe_{i,t} + DevRecipe_{i,t} \quad (12)$$

While other state transitions involve the interaction with the environment, we will describe them in Section 3.3.

The design of the reward function $r_t$ is shown in Eq.(13) with respect to the state and action described above.

$$r_t = \begin{cases} RlCost_t & \text{if } Recipemin_i \leq Recipe_{i,t} \leq Recipemax_i \\ & \text{and } DevInitRecipe_{\min} \leq DevInitRecipe_{i,t} \leq DevInitRecipe_{\max} \\ -1 & \text{else} \end{cases} \quad (13)$$

$$RlCost_t = \begin{cases} Co_{cost} \cdot (Cbase - Cmix_t) & \text{if } PLoBound_{j,t} \leq Pmix_{j,t} \leq PUpBound_{j,t} \\ -0.1 & \text{else} \end{cases} \quad (14)$$

In Eq.(13), if the recipe is within constraints, the reward function is $RlCost_t$, which represents the reward with respect to the cost of mix oil. The $RlCost_t$ is calculated by Eq.(14). If the properties of the mixed oil are between the upper and lower constraints, when $Cmix_t$ is larger, $Cost_t$ is smaller.

### 3.3 Environment description

The environment is used as a simulator for generating training data, including properties and recipes of component oils, properties and constraints of mixed oil, and properties of product oil in the tank. The performance of the proposed method is also tested in this environment. For each batch of gasoline blending, the calculation process of the environment model is shown in Table 1. For clarity of expression, only the properties of RON, density, sulfur content, olefin, and aromatics are considered. More properties can be considered by extending the state space. Catalytic gasoline, non-aromatic, Reformate, and MTBE are chosen to blend #95 gasoline in blending process 1, where the initial conditions are shown in Table 2. If the reserve of this component oil of MTBE is insufficient, another component oil C5 is used to replace it, where the initial conditions are shown in Table 3. The nominal properties and cost of the component oils are shown in Table 4 and Table 5, respectively. The prescribed standards for #95 gasoline and the initial properties range for tank bottom oil are shown in Table 6. Fluctuations will be applied to the research octane number of each component oil according to the one-year measurement data of a real-world refinery.

Table 1. Environment description

| The gasoline blending process (one batch) simulator |
|---|

Initialize $Ptank_{t0}$ according to the average sampling results of the data in column 2 in Table 6. Zero the elements of the $DevInitRecipe_{t0}$. Set the $Recipe_{t0}$ as the initial recipe for each component oil in Table 2 or Table 3. Set $Vtank_{t0} = 0.1$, which represents that the tank bottom oil accounts for 10% of the total volume of the gasoline storage tank.

For each optimization step $t$

  If gasoline storage tank is not full, then

   1. Obtain the research octane number of the component oil according to the one-year measurement data of an actual refinery (We only consider the fluctuation of RON, and the nominal values in Table 4 will be used for other properties) and form the component oil property matrix $Pcompo_{t+1}$.

   2. Get the action $DevRecipe_t$ given by the RL agent.

   3. Calculate the $Recipe_{t+1}$ and $DevInitRecipe_{t+1}$ according to the Eq.(11) and Eq.(12). Respectively.

   4. Calculate the $Cmix_{t+1} = \sum_{i=1}^{n} Ccompo_i \cdot Recipe_{i,t+1}$.

   5. Calculate the $Pmix_{t+1}$ linearly, except the RON of mixed oil $Pmix_{ron,t+1}$ is calculated by Eqs.(15)-(17).

   6. Calculate the $Vtank_{t+1} = Vtank_t + 0.02$, which represents that the mixed oil is fed into the gasoline storage tank at a constant rate.

   7. Calculate the $Ptank_{t+1}$ by linearly adding the property of the mixed oil to be input into the tank and the property of existing oil in the tank according to their volume ratio.

   8. Calculate the $PLoBound_{t+1}$ and $PUpBound_{t+1}$ according to the Eq.(8) and Eq.(9).

The Stewart model used to characterize the nonlinear mechanism of RON blending is as follows:

$$O_{blend} = \sum_{s \in S} w_s O_s \tag{15}$$

$$RON_{blend} = \frac{\sum_{s \in S} V_s D_s [RON_s + \bar{c}(O_s - O_{blend})]}{\sum_{s \in S} V_s D_s} \tag{16}$$

$$D_s = \frac{\bar{a}(O_s - O_{blend})}{1 - \exp[\bar{a}(O_s - O_{blend})]} \tag{17}$$

where the parameters $\bar{a}$ and $\bar{c}$ are set to 0.0414 and 0.01994, respectively.

Table 2. Initial conditions of blending process 1

|  | Catalytic gasoline | Non-aromatic | Reformate | MTBE |
|---|---|---|---|---|
| Initial recipe | 45 | 19 | 27 | 9 |
| Recipe change range | 5 | 5 | 5 | 5 |

Table 3. Initial conditions of blending process 2

|  | Catalytic gasoline | Non-aromatic | Reformate | C5 |
|---|---|---|---|---|
| Initial recipe | 45 | 11 | 38 | 6 |
| Recipe change range | 5 | 5 | 5 | 5 |

Table 4. Nominal properties of the component oils

|  | Catalytic gasoline | Non-aromatic | Reformate | MTBE | C5 |
|---|---|---|---|---|---|
| RON | 92.5 | 74.5 | 103 | 108 | 80 |
| Density(kg/m³) | 736 | 668 | 800 | 727 | 640 |
| Sulfur content/ppm | 15 | 0 | 0 | 0 | 0 |
| Olefin(vol%) | 30 | 2 | 0.9 | 0 | 0 |
| Aromatics(vol%) | 20 | 2.4 | 30 | 0 | 0 |

Table 5. Price of the component oils

|  | Catalytic | Non- | Reformate | MTBE | C5 |
|---|---|---|---|---|---|

|  | gasoline | aromatic |  |  |  |
|---|---|---|---|---|---|
| Cost(RMB/ton) | 4300 | 3900 | 5000 | 6000 | 4000 |

Table 6. Standards for #95 gasoline and initial properties range for tank bottom oil

|  | #95 gasoline | Tank bottom oil |
|---|---|---|
| RON | 95~ | 94.5~96.5 |
| Density(kg/m$^3$) | 720~775 | 720~775 |
| Sulfur content/ppm | 0~10 | 7~13 |
| Olefin(vol%) | 0~15 | 13~16 |
| Aromatics(vol%) | 0~35 | 18~30 |

## 4. Application to gasoline blending system

### 4.1 Determination of the optimal neural network structure

The structure of neural networks is a major influence on the performance of DRL agents. In order to find the optimal network structure, five experiments are conducted using different neural network structures. In each experiment, all neural networks adopt the same structure. Each structure adopts two hidden layers with different numbers of neurons, namely 16, 32, 64, 128, and 256. The average rewards can be seen in Fig. 4. As it can be seen, for networks with 16x16 and 32x32 neurons, the network capacity is not enough to support the RL agent in learning high performance policy. The average rewards fluctuate greatly during the training process, and do not converge to a satisfactory level. The results are quite acceptable for networks with 64x64, 128x128, and 256x256 neurons. As the number of episodes increases, the average rewards continue to increase, and are nearly stable around the 1000th episode. The maximum average reward under each network structure can be seen in Table 7. The maximum average reward under networks with 16x16 and 32x32 neurons is far exceeded by others. The maximum average reward under networks with 64x64, 128x128, and 256x256 neurons increases little with the network capacity expansion. Therefore, investigating a more complex network structure will not bring better results. Considering the complexity and performance of the network, we finally choose the 64x64 network structure, and the following experiments are carried out under this network structure.

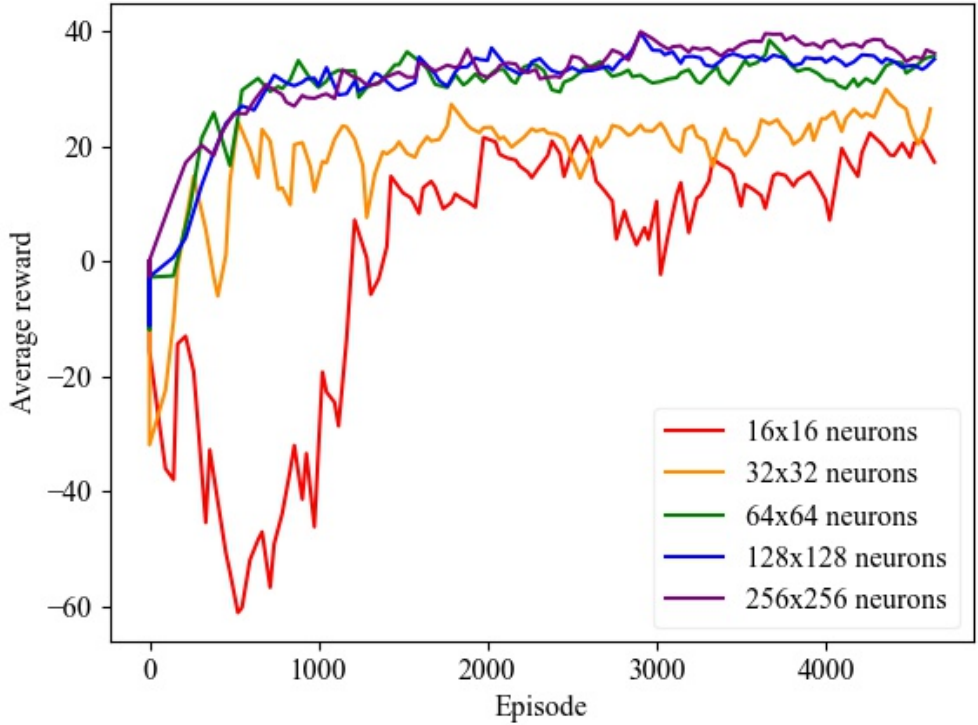

Fig. 4. Average rewards under different network structures.

Table 7. Maximum average rewards under different network structures

| No. | Neurons | Maximum average reward |
|---|---|---|
| 1 | 16×16 | 22.33 |
| 2 | 32×32 | 29.94 |
| 3 | 64×64 | 38.49 |
| 4 | 128×128 | 39.74 |
| 5 | 256×256 | 39.95 |

**4.2 Performance analysis**

In this section, the proposed DRL based method is used to solve the online gasoline blending problem. A blending effect (BE) [9] based optimization method is chosen to demonstrate the competitiveness of the proposed method. Two methods are compared under the same random seeds, so the initial properties of the oil in the tank and the change trend of component oil properties in the whole process are the same to ensure fair comparison.

We used a total of 20 random seeds for comparative experiments. As can be seen from the Table 8, the average costs of the two methods are generally the same, and the cost of DRL based policy is slightly lower. Although DRL based policy saves only 3 yuan/ton compared to BE based policy, the refinery will save millions of yuan because millions of tons gasoline need to be blended every year. In the following, a set of scenarios have been selected to show the difference in performance between the two methods.

Table 8. Average cost of the two policies

| Policy | Average cost (yuan/ton) |
| --- | --- |
| BE based policy | 4526.19 |
| DRL based policy | 4523.26 |

As shown in Fig.5, the DRL based method has a slightly higher cost of mixed oil in the initial stage of the blending process. As the blending process continues, the cost of the mixed oil continues to decrease, making its overall cost lower than the BE based method. Fig.6 shows the recipes of the two methods change in the whole process. Fig.7 shows the change of the blending properties and their allowable ranges. It can be seen from Figs.6-7 that the DRL based method first adopts the recipes of component oil with higher RON. This may be because it takes into account the property fluctuations in the process, so it adopts a conservative blending strategy at the beginning. Although this makes mixed oil more expensive, but also lowers the lower bound of RON, so that the DRL agent can use recipes with lower RON in the later stage to make up for the benefits lost in the early stage. As shown in Fig.8, the properties of the final product oil of the two methods meet the standard. But the DRL based method has a lower blending cost throughout the blending process.

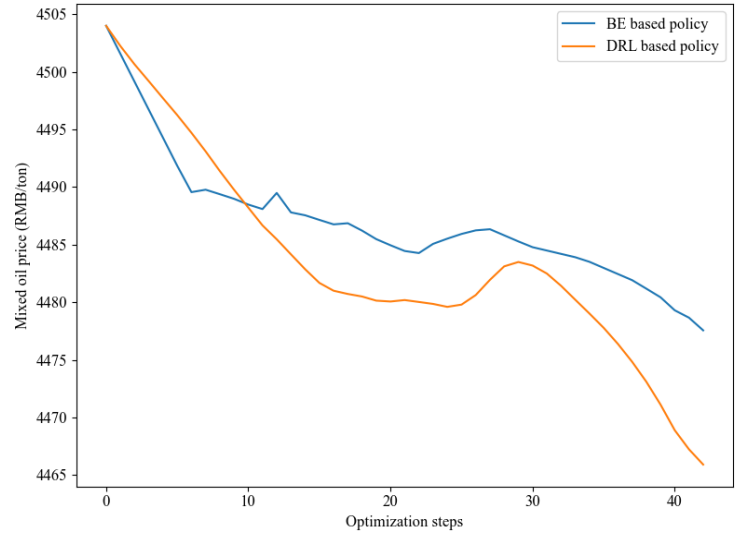

Fig. 5. Average cost of the mixed oil in blending process 1

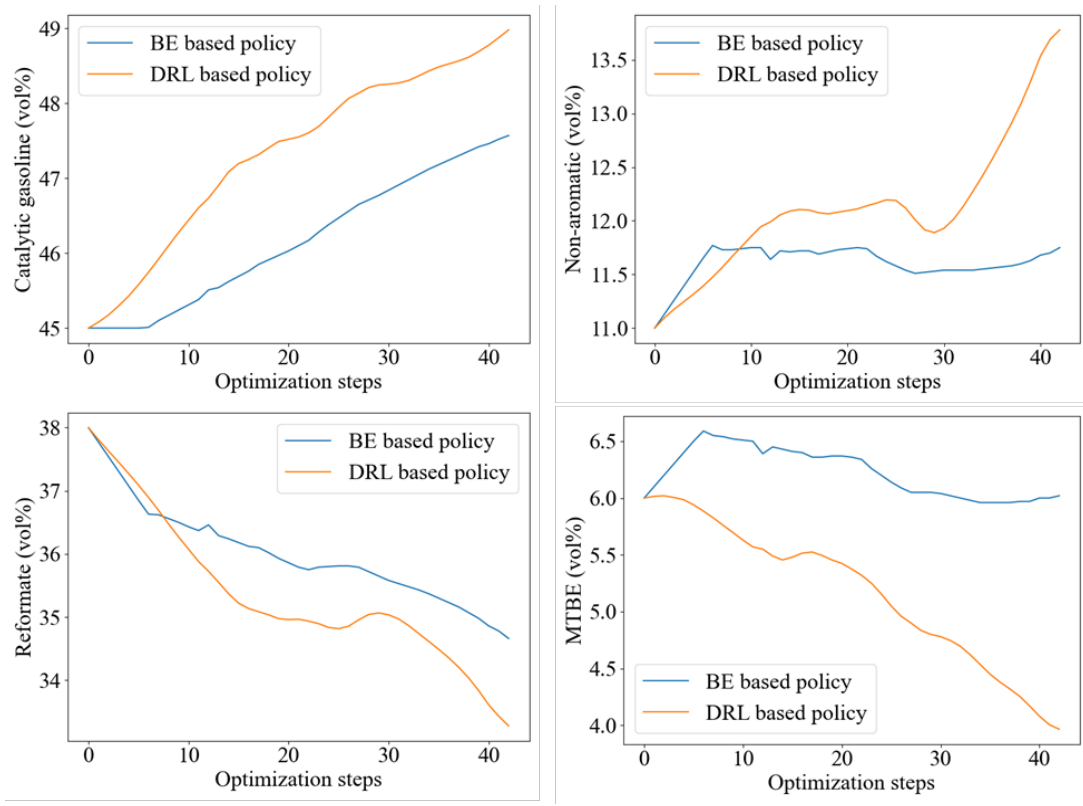

Fig. 6. Recipes in blending process 1

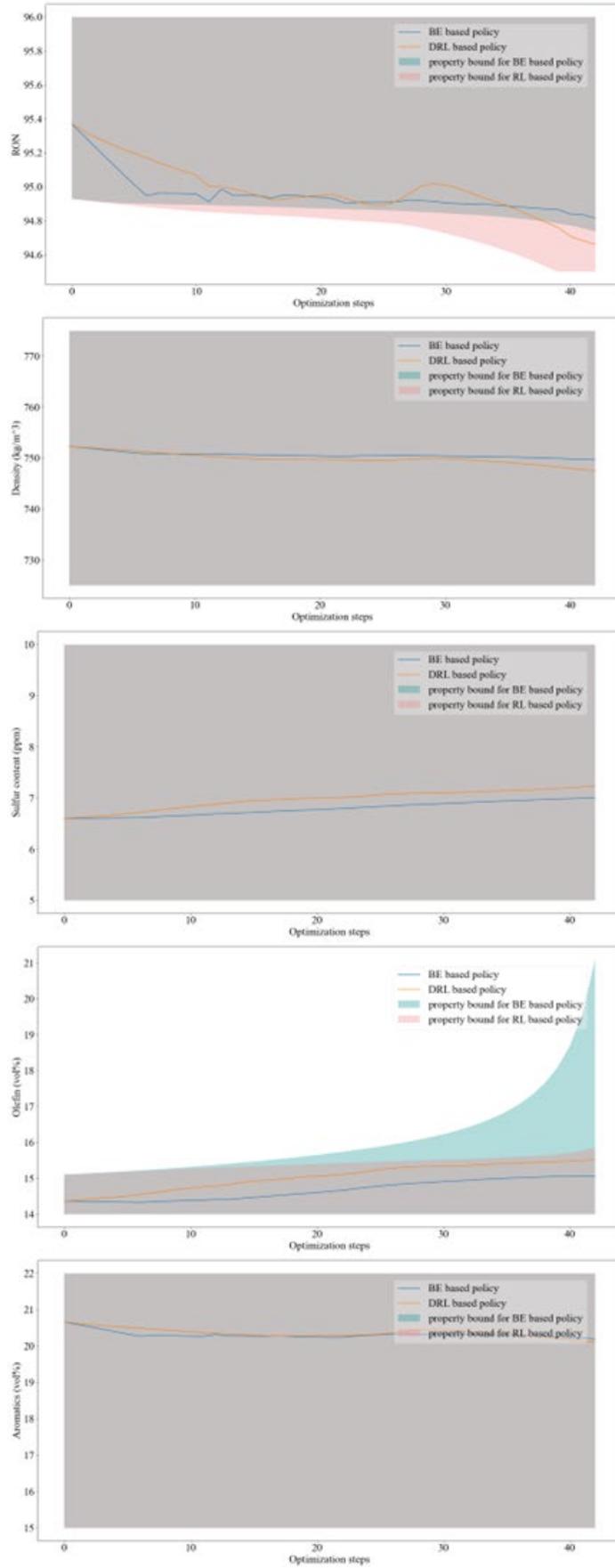

Fig. 7. Mixed oil properties and property bound in blending process

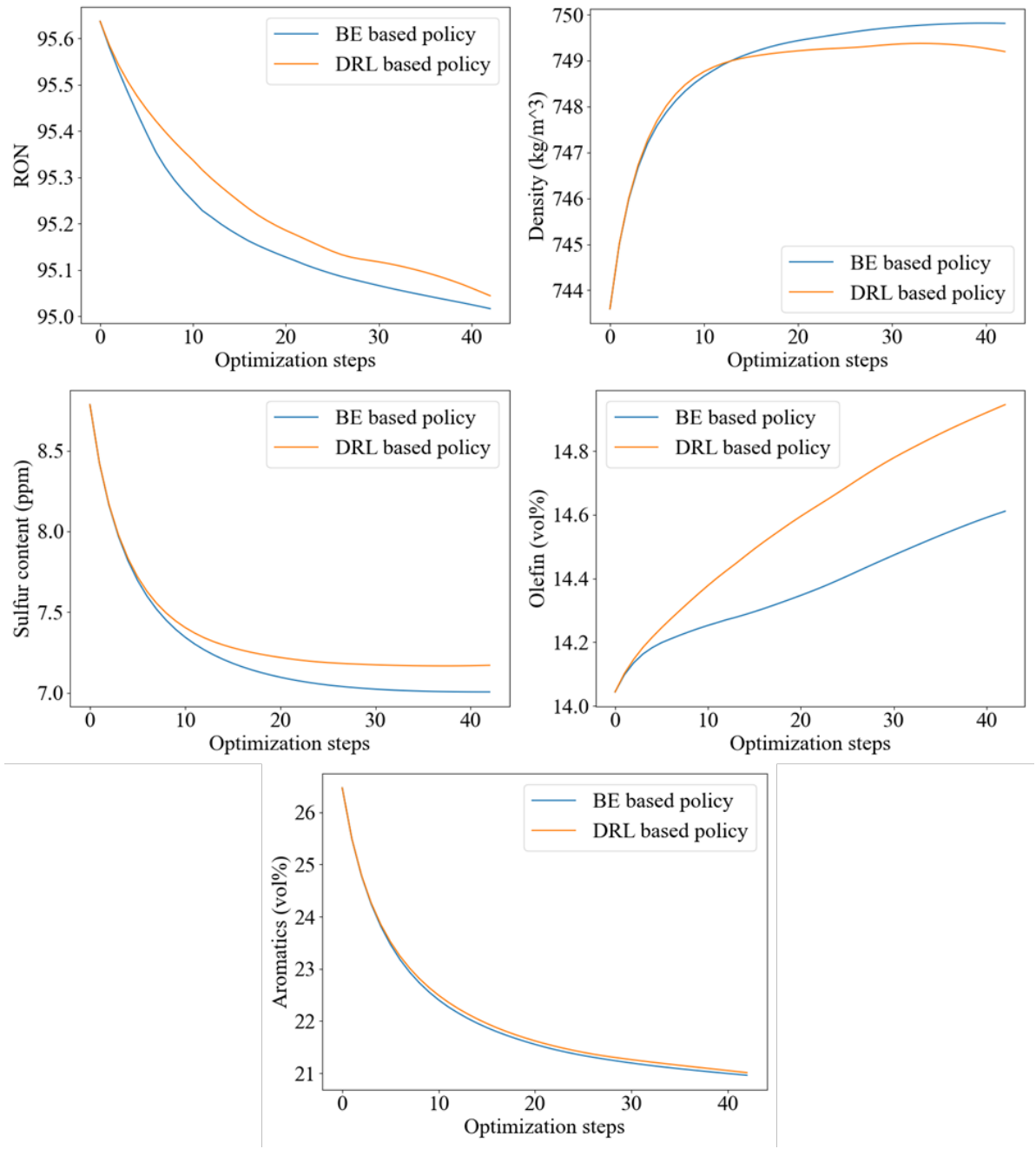

Fig. 8. Product oil properties in blending process 1

In the actual production process, there will be insufficient reserves of some component oil occasionally. As a result, switching of component oil types may occur. In view of this situation, we designed experiments to observe the performance of the two methods when the type of component oil changes in the blending process. We assume that when the blending process reaches 15th step, the reserve of component oil MTBE is insufficient, and another component oil C5 is used to replace it. As shown in Fig.9, when the type of component oil is not changed, the prices of the two are quite close. However, when the type of component oil is changed, the average cost of the BE based method rises sharply. After a certain optimization steps, the average cost of the BE based method drops to a level similar to that of the DRL based method; The method based on DRL does not produce large fluctuation at the switching time. Fig.11 shows the changes of property of the mixed oil in this situation. The change of RON of mixed oil is similar to that of the average cost. At the time of switching, the RON of the mixed oil based on the BE method increased significantly to 96. The olefin also exceeds the limit at an optimization step after switching. However, the properties of mixed oil blended based on DRL method do not fluctuate significantly. The reason for this phenomenon is that the method based on BE is linear optimization, which uses the compensation value to fit the nonlinear process of RON mixing. Therefore, when the oil type changes, it needs certain optimization steps to adjust the compensation value, so there are large fluctuations in the process after switching. The method based on DRL learns the nonlinear process of RON mixing, so the change of oil does not have a great impact on it. Observing the product oil property changing curve in Fig.12, the final product oil properties also meet the standard in this experiment. But the DRL based method shows its stability in the process.

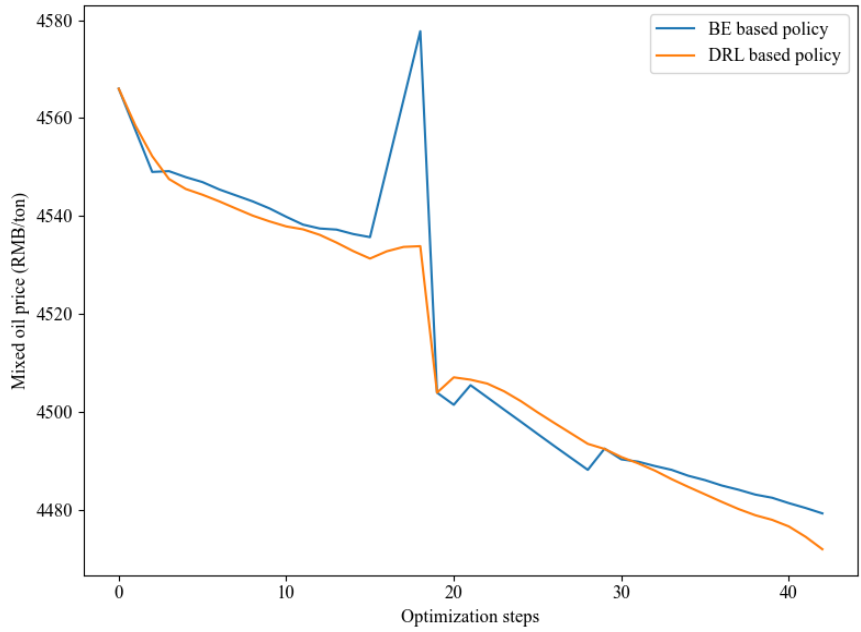

Fig. 9. Average cost of the mixed oil when changing from blending process 1 to blending process 2

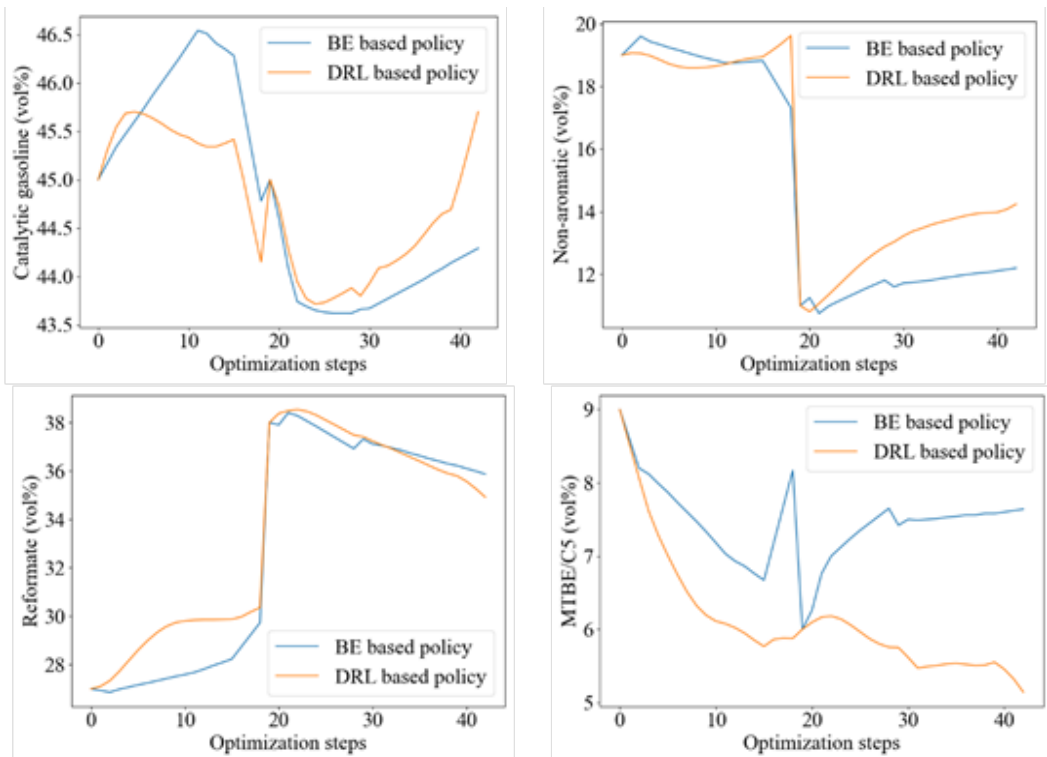

Fig. 10. Recipes when changing from blending process 1 to blending process 2

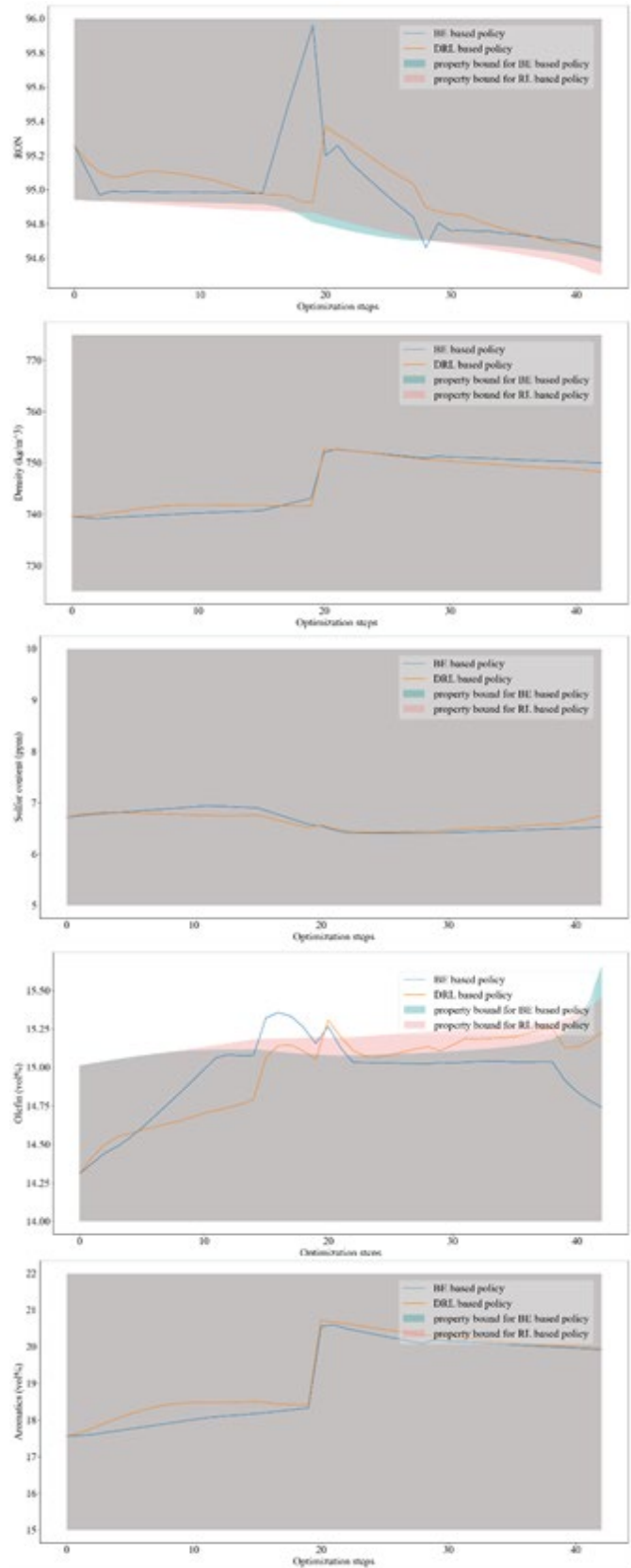

Fig. 11. Mixed oil properties and property bounds when changing from blending process 1 to blending process 2

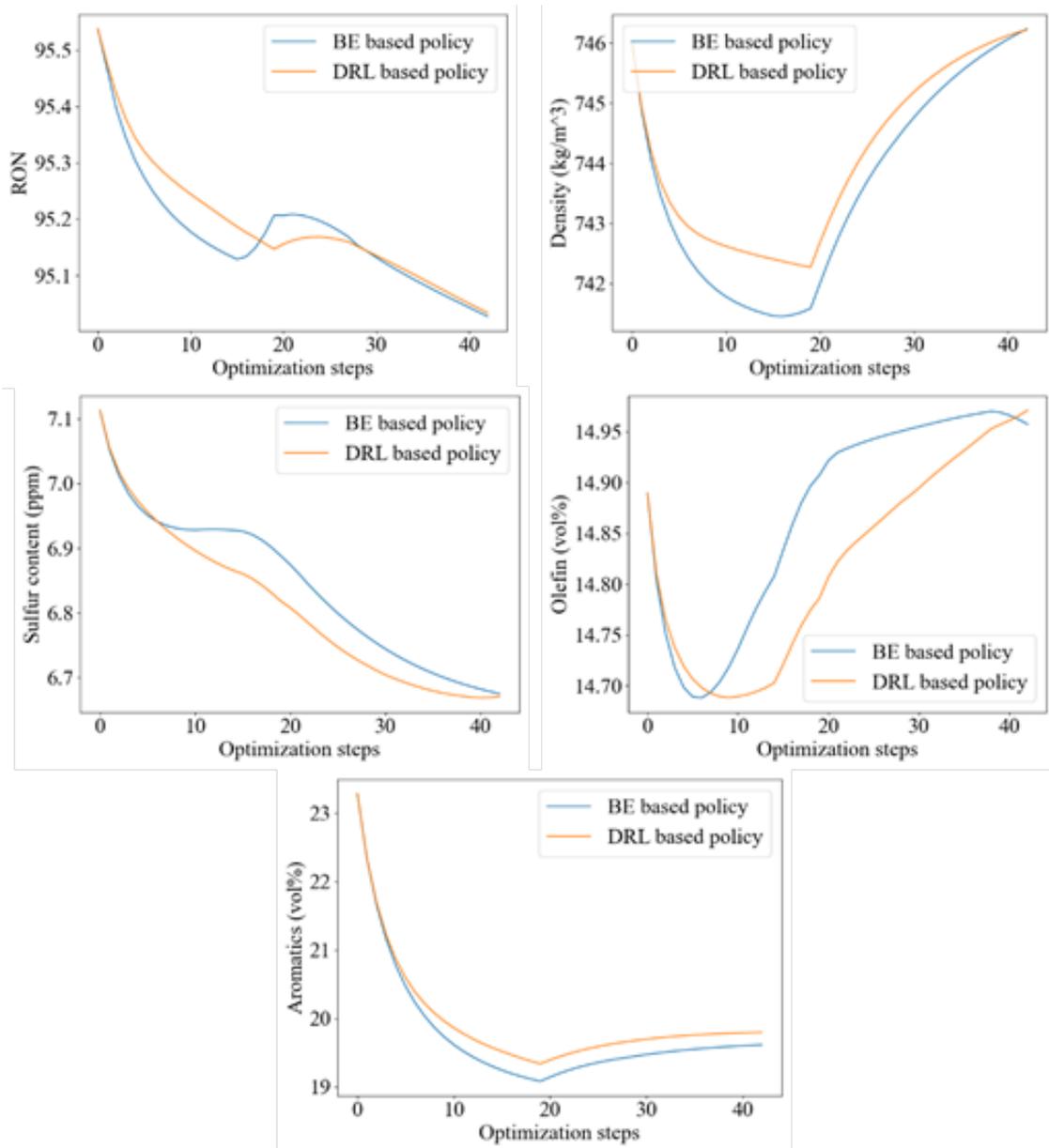

Fig. 12. Product oil properties when changing from blending process 1 to blending process 2

## 4.3 Adaptiveness to system drifting

One of the advantages of the DRL based online gasoline blending optimization is that the policy learned by the agent is continuously updated and follows the drift of the system. Specifically, as the old data in the data buffer is constantly replaced by the new data during the interaction, the value network and soft Q network updates the policy by tracking this change.

In the actual process, the model mismatch caused by the fluctuation of component oil properties or plant/model parametric mismatch may greatly affect the performance of traditional model-based optimization methods. In the proposed DRL based method, when the system is drifting, the process of replacing old data with new data make the data buffer always match the actual environment. This further enables the policy to automatically follow the system drift.

In this experiment, after the policy is stabilized at 2200th episode, we change the parameter in Eqs.(15)-(17) from (0.0414, 0.01994) to (0.414, 0.01994). It can be observed from Fig.13 that the average reward continues to rise and stabilize when training with the original parameters. And when the parameters are changed, the average reward continues to rise and stabilizes under the new parameters. Due to the change of parameters in the blending mechanism model, blending gasoline of the same quality requires a lower cost of mixed oil, so the reward for the new steady state is higher.

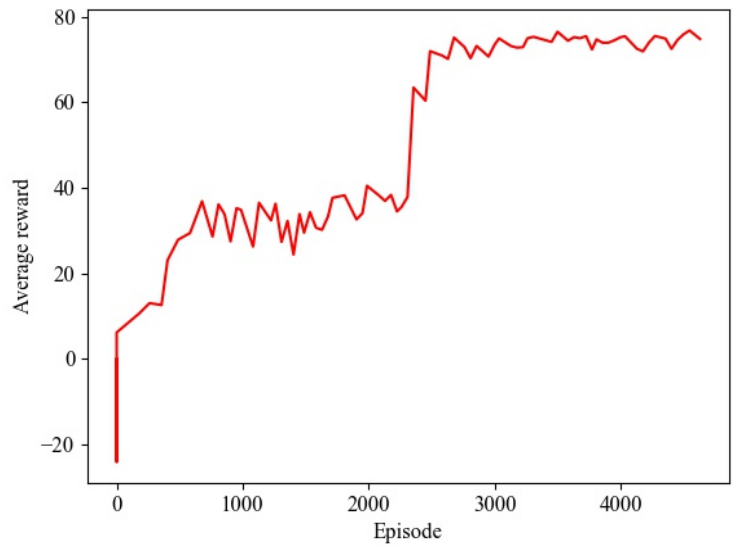

Fig. 13. Average rewards when system drifting.

## 5. Conclusions

In this paper, a gasoline blending on-line optimization method based on DRL algorithm is proposed to avoid complex nonlinearity, fluctuation of blending component oil attributes and possible blending model mismatch in the process of gasoline blending. Aiming at an actual gasoline blending system, its MDP expression and simulation environment is established, and solved by DRL algorithm SAC. The results show that the proposed method has better economic performance and it is more robust under property fluctuation and blending component oil switching. In addition, this method can adapt to drift and maintain performance without manual intervention. It is proved that the method is effective in the online optimization of gasoline blending.

## Acknowledgments

The authors acknowledge the supports from National Natural Science Foundation of China (Basic Science Center Program: 61988101), International (Regional) Cooperation and Exchange Project (61720106008), and National Natural Science Foundation of China (62073142).

## Nomenclature

| | |
|---|---|
| $Pcompo_t$ | Property matrix of all component oils at time $t$ |
| $Pcompo_{i,t}$ | Property vector of the $i$-th component oil |
| $Pcompo_{i,j,t}$ | The $j$-th property of the $i$-th component oil at time $t$ |
| $Recipe_t$ | Recipe vector of all component oils at time $t$ |
| $Recipe_{i,t}$ | Recipe of the $i$-th component oil at time $t$ |
| $DevInitRecipe_t$ | Recipe deviation (from initial recipe) vector of all component oils at time $t$ |
| $DevInitRecipe_{i,t}$ | Recipe deviation (from initial recipe) of the $i$-th component oil at time $t$ |

| Symbol | Description |
|---|---|
| $Cmix_t$ | Price per ton of mixed oil at time $t$ |
| $Pmix_t$ | Property vector of mixed oil at time $t$ |
| $Pmix_{j,t}$ | The $j$-th property of mixed oil at time $t$ |
| $Ptank_t$ | Property vector of product oil at time $t$ |
| $Ptank_{j,t}$ | The $j$-th property of product oil at time $t$ |
| $Vtank_t$ | Volume of product oil at time $t$ |
| $V_{target}$ | Target volume of product oil |
| $PLoBound_t$ | Property lower bound vector of mixed oil at time $t$ |
| $PLoBound_{j,t}$ | The $j$-th property lower bound of mixed oil at time $t$ |
| $PUpBound_t$ | Property upper bound vector of mixed oil at time $t$ |
| $PUpBound_{j,t}$ | The $j$-th property upper bound of mixed oil at time $t$ |
| $PLoPO_j$ | The $j$-th property lower bound of product oil |
| $PUpPO_j$ | The $j$-th property upper bound of product oil |
| $DevRecipe_t$ | Recipe change vector at time $t$ |
| $DevRecipe_{i,t}$ | Recipe change of the $i$-th component oil at time $t$ |
| $Cbase$ | Price base for calculating the reward |
| $Co_{volume}$ | Constant before volume of product oil |